\documentclass[nofootinbib,prd,twocolumn,showpacs,showkeys,preprintnumbers]{revtex4-1}
\usepackage{hyperref,amssymb,amsmath,mathrsfs,bm,graphicx}
\usepackage[utf8]{inputenc}
\usepackage{enumitem}
\input{epsf}

\begin{document}
\title {Complexity and symmetry: The asymmetry factor approach }
\author{L. Herrera}
\email{lherrera@usal.es}
\affiliation{Instituto Universitario de F\'isica
Fundamental y Matem\'aticas, Universidad de Salamanca, Salamanca 37007, Spain.}
.
\date{\today}
\begin{abstract}
We study the influence of the instantaneous  appearance  of a time-like Killing vector (TKV) in self-gravitating  fluid spheres during their evolution.  For doing so we resort to the concept of asymmetry factor which  allows the existence of any specific type of symmetry (e.g. a TKV), exclusively,   for a given value of the time-like coordinate. We consider adiabatic and dissipative fluids. The analysis of different relevant physical variables in this process shows the imprints of the emergence of TKV at some point of the evolution. Particular attention is paid to the relationship appearing between the TKV, and the complexity factor describing the degree  of complexity of the fluid, bringing out for the first time a  link between complexity and symmetry.
 \end{abstract}
\date{\today}
\maketitle

\section{Introduction}
In a recent paper \cite{1} we have introduced the concept of asymmetry factor, which is a scalar function that allows  to switch on (off) a given symmetry for any specific value  of the time-like coordinate. In \cite{1} we considered the case of a conformal Killing vector, and we looked for the imprints that the instantaneous appearance of such  a symmetry produces on the evolution of the fluid. 

In  this work we want to apply a similar scheme to the case where the involved symmetry corresponds to a time-like Killing vector field, focusing on the possible link between the complexity as measured by the complexity factor \cite{ps1,ps2}, and the symmetry under consideration. 

The idea of complexity in different branches of science  has a long and a venerable history (see \cite{ts} and references therein). Some years ago important advances on  this issue were achieved by Lopez-Ruiz and collaborators within  the context of statistical physics \cite{3,3bisbis,4,4bis}. This approach was later on employed by some authors to define complexity for self-gravitating systems (see \cite{10}  and references therein). However such a definition suffers from some flaws, which  motivated the search for an alternative definition of complexity for self-gravitating systems. This search led to the definition proposed in \cite{ps1,ps2}.

Thus we shall consider  spherically symmetric fluid distributions which at some point of their  evolution admit a TKV (the particular  case of geodesic fluids has been studied in \cite{ul}).  The (dis)appearance of the TKV is controlled by a tensor defined through the asymmetry factor. As we will  see below, the sudden appearance  of the TKV, even during  an infinitesimal small interval of time,  has a remarkable influence on all physical variables (in particular on the complexity factor).

We shall consider two different models. The first one is adiabatic (non-dissipative).  Adiabatic  models, in general,  are defined up  to two functions of the radial coordinate. A specific choice of these functions provides us  with our first model.   The evolution of such a system is analyzed in detail  assuming that for  some value of the time-like coordinate, the space-time admits a TKV which vanishes immediately afterward. 

Next, we shall consider dissipative fluids.  In this case it can be shown that the fluid cannot  be shear-free, and the system is defined up to three arbitrary functions of $r$. Our second model is obtained from a specific choice of these three functions.

In both cases, all the physical variables are  analyzed in detail. In the adiabatic case particular attention is paid to the effects of the sudden appearance of the TKV on the behavior of the complexity factor and the anisotropic factor, whereas in the dissipative case we have focused on the behavior of the temperature,   the redshift and the complexity factor.

The manuscript is organized  as follows: the method and the variables required for describing   the evolution of a spherically symmetric fluid distribution  are  presented  in the next section.  In Section III we introduce the concept of asymmetry factor for a time-like isometry, and express the metric variables in terms of such a factor.
The non-dissipative and the dissipative cases  are  analyzed in section IV, and   section V respectively.  A  discussion on  the obtained results   is  presented in section VI .
\section{Fundamental variables  and relevant equations}
Let us briefly present   definitions and main equations required for describing the structure and evolution of spherically symmetric dissipative fluids. For detail see  \cite{ps2} and references therein.

We consider  spherically symmetric distributions  of evolving
fluids, which  are  bounded by a spherical surface $\Sigma$. The fluid is
assumed to be locally anisotropic (principal stresses unequal) and undergoing dissipation in the
form of heat flow (diffusion approximation).

Choosing comoving coordinates, the general
interior metric can be written as
\begin{equation}
ds^2=-A^2dt^2+B^2dr^2+R^2(d\theta^2+\sin^2\theta d\phi^2),
\label{1}
\end{equation}
where $A$, $B$ and $R$ are functions of $t$ and $r$ and are assumed
positive. We number the coordinates $x^0=t$, $x^1=r$, $x^2=\theta$
and $x^3=\phi$, where $r$ is a radial comoving coordinate.  Observe that $A$ and $B$ are dimensionless, whereas $R$ has the same dimension as $r$.

The energy--momentum tensor $T_{\alpha\beta}$ of the fluid distribution
has the form
\begin{eqnarray}
T_{\alpha\beta}&=&(\mu +
P_{\perp})V_{\alpha}V_{\beta}+P_{\perp}g_{\alpha\beta}+(P_r-P_{\perp})K_{
\alpha}K_{\beta}\nonumber \\&+&q_{\alpha}V_{\beta}+V_{\alpha}q_{\beta}
, \label{3}
\end{eqnarray}
where $\mu$ is the energy density, $P_r$ the radial pressure,
$P_{\perp}$ the tangential pressure, $q^{\alpha}$ the heat flux, $V^{\alpha}$ the four-velocity of the fluid,
and $K^{\alpha}$ a unit four-vector along the radial direction. These quantities
satisfy
\begin{eqnarray}
V^{\alpha}V_{\alpha}=-1, \;\; V^{\alpha}q_{\alpha}=0, \;\; K^{\alpha}K_{\alpha}=1,\;\;
K^{\alpha}V_{\alpha}=0.
\end{eqnarray}

It will be convenient to express the  energy momentum tensor  (\ref{3})  in the equivalent (canonical) form
\begin{equation}
T_{\alpha \beta} = {\mu} V_\alpha V_\beta + P h_{\alpha \beta} + \Pi_{\alpha \beta} +
q \left(V_\alpha K_\beta + K_\alpha V_\beta\right) \label{Tab}
\end{equation}
with
$$ P=\frac{P_{r}+2P_{\bot}}{3}, \qquad h_{\alpha \beta}=g_{\alpha \beta}+V_\alpha V_\beta,$$

$$\Pi_{\alpha \beta}=\Pi\left(K_\alpha K_\beta - \frac{1}{3} h_{\alpha \beta}\right), \qquad \Pi=P_{r}-P_{\bot}.$$

In the above $P$ is the isotropic pressure and $ h_{\alpha \beta}$ is the projector on the hypersurface orthogonal to the four-velocity.

Since we are considering comoving observers, we have
\begin{eqnarray}
V^{\alpha}&=&A^{-1}\delta_0^{\alpha}, \;\;
q^{\alpha}=qK^{\alpha}, \;\;
K^{\alpha}=B^{-1}\delta^{\alpha}_1,
\end{eqnarray}
where $q\equiv\sqrt{q^\mu q_\mu}$ is a function of $t$ and $r$, measuring the strength of the dissipated energy.

It is worth noticing that we do not explicitly add bulk or shear viscosity to the system because they
can be trivially absorbed into the radial and tangential pressures, $P_r$ and
$P_{\perp}$, of the collapsing fluid (in $\Pi$). Also we do not explicitly  introduce  dissipation in the free streaming approximation since it can be absorbed in $\mu, P_r$ and $q$.

The Einstein equations are given by
\begin{equation}
G_{\alpha\beta}=8\pi T_{\alpha\beta},
\label{2}
\end{equation}
and its non zero components
with (\ref{1}) and (\ref{3}) 
become

\begin{eqnarray}
8\pi T_{00}&=&8\pi  \mu A^2
=\left(2\frac{\dot{B}}{B}+\frac{\dot{R}}{R}\right)\frac{\dot{R}}{R}\nonumber \\
&-&\left(\frac{A}{B}\right)^2\left[2\frac{R^{\prime\prime}}{R}+\left(\frac{R^{\prime}}{R}\right)^2
-2\frac{B^{\prime}}{B}\frac{R^{\prime}}{R}-\left(\frac{B}{R}\right)^2\right],\nonumber \\
\label{12} 
\end{eqnarray}
\begin{eqnarray}
8\pi T_{01}=-8\pi qAB
=-2\left(\frac{{\dot R}^{\prime}}{R}
-\frac{\dot B}{B}\frac{R^{\prime}}{R}-\frac{\dot
R}{R}\frac{A^{\prime}}{A}\right),
\label{13} 
\end{eqnarray}
\begin{eqnarray}
8\pi T_{11}=8\pi P_r B^2
&=-&\left(\frac{B}{A}\right)^2\left[2\frac{\ddot{R}}{R}-\left(2\frac{\dot A}{A}-\frac{\dot{R}}{R}\right)
\frac{\dot R}{R}\right]\nonumber \\
&+&\left(2\frac{A^{\prime}}{A}+\frac{R^{\prime}}{R}\right)\frac{R^{\prime}}{R}-\left(\frac{B}{R}\right)^2,
\label{14} 
\end{eqnarray}
\begin{eqnarray}
8\pi T_{22}=\frac{8\pi}{\sin^2\theta}T_{33}=8\pi P_{\perp}R^2
&=-&\left(\frac{R}{A}\right)^2\left[\frac{\ddot{B}}{B}+\frac{\ddot{R}}{R}
-\frac{\dot{A}}{A}\left(\frac{\dot{B}}{B}+\frac{\dot{R}}{R}\right)\right. \nonumber \\ \left.
+\frac{\dot{B}}{B}\frac{\dot{R}}{R}\right]
&+&\left(\frac{R}{B}\right)^2\left[\frac{A^{\prime\prime}}{A}
+\frac{R^{\prime\prime}}{R}-\frac{A^{\prime}}{A}\frac{B^{\prime}}{B}\right. \nonumber \\ \left.
+\left(\frac{A^{\prime}}{A}-\frac{B^{\prime}}{B}\right)\frac{R^{\prime}}{R}\right].\label{15}
\end{eqnarray}

The component (\ref{13}) can be rewritten with (\ref{5c1}) and
(\ref{5b1}) as
\begin{equation}
4\pi qB=\frac{1}{3}(\Theta-\sigma)^{\prime}
-\sigma\frac{R^{\prime}}{R}.\label{17a}
\end{equation}

The acceleration $a_{\alpha}$ and the expansion $\Theta$ of the fluid are
given by
\begin{equation}
a_{\alpha}=V_{\alpha ;\beta}V^{\beta}, \;\;
\Theta={V^{\alpha}}_{;\alpha}. \label{4b}
\end{equation}
and its  shear $\sigma_{\alpha\beta}$ by
\begin{equation}
\sigma_{\alpha\beta}=V_{(\alpha
;\beta)}+a_{(\alpha}V_{\beta)}-\frac{1}{3}\Theta h_{\alpha\beta}.
\label{4a}
\end{equation}

From  (\ref{4b}) we have for the  four--acceleration and its scalar $a$,
\begin{equation}
a_\alpha=a K_\alpha, \;\; a=\frac{A^{\prime}}{AB}, \label{5c}
\end{equation}
and for the expansion
\begin{equation}
\Theta=\frac{1}{A}\left(\frac{\dot{B}}{B}+2\frac{\dot{R}}{R}\right),
\label{5c1}
\end{equation}
where the  prime stands for $r$
differentiation and the dot stands for differentiation with respect to $t$.

The shear tensor (\ref{4a}) has only one independent non zero component, and may be written as 
\begin{equation}
\sigma_{\alpha \beta}=\sigma \left(K_\alpha K_\beta-\frac{h_{\alpha \beta}}{3}\right),
\label{defs}
\end{equation}
with 
\begin{equation}
\sigma=\frac{1}{A}\left(\frac{\dot{B}}{B}-\frac{\dot{R}}{R}\right).\label{5b1}
\end{equation}

Next, the mass function $m(t,r)$  reads
\begin{equation}
m=\frac{R^3}{2}{R_{23}}^{23}
=\frac{R}{2}\left[\left(\frac{\dot R}{A}\right)^2-\left(\frac{R^{\prime}}{B}\right)^2+1\right].
 \label{17masa}
\end{equation}

Introducing the proper time derivative $D_T$
given by
\begin{equation}
D_T=\frac{1}{A}\frac{\partial}{\partial t}, \label{16}
\end{equation}
we can define the velocity $U$ of the 
fluid  as the variation of the areal radius with respect to proper time, i.e.
\begin{equation}
U=D_TR , \label{19}
\end{equation}
where $R$ defines the areal radius of a spherical surface inside the fluid distribution (as
measured from its area).

Then (\ref{17masa}) can be rewritten as
\begin{equation}
E \equiv \frac{R^{\prime}}{B}=\left(1+U^2-\frac{2m}{R}\right)^{1/2}.
\label{20x}
\end{equation}
Using (\ref{20x}) we can express (\ref{17a}) as
\begin{equation}
4\pi q=E\left[\frac{1}{3}D_R(\Theta-\sigma)
-\frac{\sigma}{R}\right],\label{21a}
\end{equation}
where   $D_R$ denotes the proper radial derivative,
\begin{equation}
D_R=\frac{1}{R^{\prime}}\frac{\partial}{\partial r}.\label{23a}
\end{equation}
\\

\subsection{The complexity factor}
The complexity factor is a  scalar function intended to describe the degree of complexity of a fluid distribution \cite{ps1,ps2}. It is identified with one of the structure scalars \cite{sc}, more specifically with the scalar defining the trace-free part of the electric Riemann tensor.

Thus, the tensor $Y_{\alpha \beta}$  defining the electric part of the Riemann tensor   given  by 
\begin{equation}
Y_{\alpha \beta}=R_{\alpha \gamma \beta \delta}V^\gamma V^\delta,
\label{electric}
\end{equation}
may be written as 

\begin{eqnarray}
Y_{\alpha\beta}=\frac{1}{3}Y_T h_{\alpha
\beta}+Y_{TF}(K_{\alpha} K_{\beta}-\frac{1}{3}h_{\alpha
\beta})\label{electric'}.
\end{eqnarray}

The scalar $Y_{TF}$ describing the trace-free part of $Y_{\alpha\beta}$ is identified with the complexity factor and may be written as (see \cite{sc, ps1, ps2} for details)
\begin{equation}
Y_{TF}= -8\pi\Pi +\frac{4\pi}{R^3}\int^r_0{R^3\left(D_R {\mu}-3{q}\frac{U}{RE}\right)R^\prime dr}.
\label{Y}
\end{equation}

Thus the scalar $Y_{TF}$ may be expressed  in terms of the anisotropy of pressure, the density inhomogeneity and  the dissipative variables.

In terms of the metric functions the scalar $Y_{TF}$ reads

\begin{eqnarray}
Y_{TF}= \frac{1}{A^2}\left[\frac{\ddot R}{R} - \frac{\ddot B}{B} + \frac{\dot A}{A}\left(\frac{\dot B}{B} - \frac{\dot R}{R}\right)\right]+ \nonumber \\ \frac{1}{ B^2} \left[\frac{A^{\prime\prime}}{A} -\frac{A^{\prime}}{A}\left(\frac{B^{\prime}}{B}+\frac{R^{\prime}}{R}\right)\right] .
\label{itfm}
\end{eqnarray}

\subsection{The exterior spacetime and junction conditions}
When considering  bounded fluid distributions  we have to satisfy the junction (Darmois) conditions if we wish to avoid the presence of thin shells. 

Thus, outside $\Sigma$ we assume that we have the Vaidya
spacetime (i.e.\ we assume all outgoing radiation is massless),
described by
\begin{equation}
ds^2=-\left[1-\frac{2M(v)}{\rho}\right]dv^2-2d\rho dv+\rho^2(d\theta^2
+\sin^2\theta
d\phi^2) \label{1int},
\end{equation}
where $M(v)$  denotes the total mass,
 $v$ is the retarded time, and $\rho$ is a null coordinate.

The matching of the full non--adiabatic sphere  (including viscosity) to
the Vaidya spacetime, on the surface $r=r_{\Sigma}=$ constant, was discussed in
\cite{chan1}.

From the continuity of the first  differential form it follows (see \cite{chan1} for details), 
\begin{equation}
A dt\stackrel{\Sigma}{=}d\tau, \label{junction1f}
\end{equation}
\begin{equation}
R\stackrel{\Sigma}{=}\rho(v), \label{junction1f2}
\end{equation}
and 
 \begin{equation}
\left(\frac{dv}{d\tau}\right)^{-2}\stackrel{\Sigma}{=}\left(1-\frac{2m}{\rho}+2\frac{d\rho}{dv}\right), \label{junction1f3}
\end{equation}
where $\tau$ denotes the proper time measured on $\Sigma$.

The continuity of the second differential form produces
\begin{equation}
m(t,r)\stackrel{\Sigma}{=}M(v), \label{junction1}
\end{equation}
and

\begin{eqnarray}
&&2\left(\frac{{\dot R}^{\prime}}{R}-\frac{\dot B}{B}\frac{R^{\prime}}{R}-\frac{\dot R}{R}\frac{A^{\prime}}{A}\right)
\stackrel{\Sigma}{=}-\frac{B}{A}\left[2\frac{\ddot R}{R}
-\left(2\frac{\dot A}{A}
-\frac{\dot R}{R}\right)\frac{\dot R}{R}\right]\nonumber \\&+&\frac{A}{B}\left[\left(2\frac{A^{\prime}}{A}
+\frac{R^{\prime}}{R}\right)\frac{R^{\prime}}{R}-\left(\frac{B}{R}\right)^2\right],
\label{j2}
\end{eqnarray}

where $\stackrel{\Sigma}{=}$ means that both sides of the equation
are evaluated on $\Sigma$ (observe a misprint in eq.(40) in \cite{chan1} and a slight difference in notation).

Comparing (\ref{j2}) with  (\ref{13}) and (\ref{14}) one obtains
\begin{equation}
q\stackrel{\Sigma}{=}P_r.\label{j3}
\end{equation}
Thus   the matching of
(\ref{1})  and (\ref{1int}) on $\Sigma$ implies (\ref{junction1}) and  (\ref{j3}).

Also, we have
\begin{equation}
q\stackrel{\Sigma}{=}\frac{L}{4\pi \rho^2}, \label{20lum}
\end{equation}
where $L_\Sigma$ denotes   the total luminosity of the  sphere as measured on its surface and is given by
\begin{equation}
L \stackrel{\Sigma}{=}L_{\infty}\left(1-\frac{2m}{\rho}+2\frac{d\rho}{dv}\right)^{-1}, \label{14a}
\end{equation}
and where
\begin{equation}
L_{\infty} =-\frac{dM}{dv}\stackrel{\Sigma}{=} -\left[\frac{dm}{dt}\frac{dt}{d\tau}\left(\frac{dv}{d\tau}\right)^{-1}\right],\label{14b}
\end{equation}
is the total luminosity measured by an observer at rest at infinity, which is defined by 
\begin{equation}
L_\infty=4\pi q_\Sigma\left[R^2\left(\frac{R^\prime}{B}+\frac{\dot R}{A}\right)^2 \right]_\Sigma.
\label{lum}
\end{equation}

The boundary redshift $z_\Sigma$ is given by
\begin{equation}
\frac{dv}{d\tau}\stackrel{\Sigma}{=}1+z,
\label{15b}
\end{equation}
with
\begin{equation}
\frac{dv}{d\tau}\stackrel{\Sigma}{=}\left(\frac{R^{\prime}}{B}+\frac{\dot R}{A}\right)^{-1}.
\label{16b}
\end{equation}
Therefore the time of formation of an apparent horizon is given by
\begin{equation}
\left(\frac{R^{\prime}}{B}+\frac{\dot R}{A}\right)\stackrel{\Sigma}{=}E+U\stackrel{\Sigma}{=}0.
\label{17b}
\end{equation}

When Darmois conditions are not satisfied, the boundary surface is a thin shell and we have to assume the Israel conditions \cite{isjc}.

\subsection{The transport equation}
In order to obtain an expression for the temperature we need a transport equation. Assuming a causal dissipative theory (e.g.the Israel-- Stewart theory \cite{19nt, 20nt, 21nt} ) the transport equation for the heat flux reads
\begin{eqnarray}
\tilde \tau h^{\alpha \beta}V^\gamma q_{\beta;\gamma}&+&q^\alpha=-\kappa h^{\alpha \beta}\left(T_{,\beta}+Ta_\beta\right)\nonumber \\&-&\frac{1}{2}\kappa T^2 \left(\frac{\tilde \tau V^\beta}{\kappa T^2}\right)_{;\beta} q^\alpha.
\label{tre}
\end{eqnarray}
where $\kappa$ denotes the thermal conductivity, and $T$ and $\tilde \tau$ denote temperature and relaxation time respectively. 

In the spherically symmetric case under consideration, the transport equation has only one independent component which may be obtained from (\ref{tre}) by contracting with the unit spacelike vector $K^\alpha$, we get
\begin{equation}
\tilde \tau V^\alpha  q_{,\alpha}+q=-\kappa \left(K^\alpha T_{,\alpha}+T a\right)-\frac{1}{2}\kappa T^2\left(\frac{\tilde \tau V^\alpha}{\kappa T^2}\right)_{;\alpha} q.
\label{5}
\end{equation}

Sometimes it is possible to simplify the equation above, in the so called truncated  transport equation, when the last term in (\ref{tre}) may be neglected \cite{t8}, producing 
\begin{equation}
\tilde \tau V^\alpha  q_{,\alpha}+q=-\kappa \left(K^\alpha T_{,\alpha}+T a\right).
\label{5trun}
\end{equation}

\section{The asymmetry factor for isometries}
Spacetimes admitting a TKV, satisfy  the equation
\begin{equation}
\mathcal{L}_\chi g_{\alpha \beta} =0, 
\label{1cmh}
\end{equation}
 where $\mathcal{L}_\chi$ denotes the Lie derivative with respect to the vector field ${\bold \chi}$, which in our case is time-like, and  has the form
\begin{equation}
\mathbf{\chi }= \partial _{\mathbf{t}}.
\label{2cmh}
\end{equation}

 In this work we shall consider fluid distributions which in general do not admit  a TKV, except at a given moment of their evolution. By this we mean the momentarily vanishing of $\mathcal{L}_\chi g_{\alpha \beta}$ in the zero-shifting comoving slicing.
 
 Thus let us consider 

\begin{equation}
\mathcal{L}_\chi g_{\alpha \beta}=H_{\alpha \beta},
\label{mod1n}
\end{equation}
implying
\begin{equation}
\chi^\rho \partial_\rho g_{\alpha \beta} +g_{\alpha \rho}\partial_\beta \chi^\rho+g_{\beta \rho}\partial_\alpha \chi^\rho=H_{\alpha \beta},
\label{mod11nn}
\end{equation}
where $H_{\alpha \beta}$ is (so far) an arbitrary symmetric tensor. Obviously  the vector $\chi$ defines a TKV  iff  $H_{\alpha \beta}=0$.

Using (\ref{mod11nn}) one obtains for (\ref{1})
$H_{01}=H_{02}= H_{03}=H_{13}=H_{12}=  H_{23}=0$.

Next, let us  assume for  the tensor $H_{\alpha \beta}$ the form
\begin{equation}
H_{\alpha \beta}=\alpha H_{T} \frac{g_{\alpha \beta}}{4}+\alpha H_{TF}  \left(K_\alpha K_\beta-\frac{1}{3}h_{\alpha \beta}\right),
\label{tkv1}
\end{equation}
where $H_T $ and $H_{TF}$ are two scalar dimensionless functions depending only on $t$, which define the tensor $H_{\alpha \beta}$. These scalar functions are the asymmetry factors of our problem (below we shall see that in our case one may put one of them to zero, reducing the number of asymmetry factors to one. It describes the deviation of the  space-time under consideration from the space-time admitting a TKV.  It is worth noticing that it is not the most general  form of tensor  $H_{\alpha \beta}$, it was chosen  so for simplicity.

Then, from the above the following equations are obtained
\begin{equation}
\dot A=\frac{\alpha H_{T}}{8} A,
\label{nd1}
\end{equation}
\begin{equation}
\dot B=\frac{\alpha B H_{T}}{8}+\frac{\alpha H_{TF}, B}{3}
\label{nd2}
\end{equation}
\begin{equation}
\dot R=\frac{\alpha RH_{T}}{8}-\frac{\alpha H_{TF} R}{6},
\label{nd3}
\end{equation}
where $\alpha$ is a unit constant with dimensions $\frac{1}{length}$.

Integrating (\ref{nd1}) we obtain
\begin{equation}
A=f(r) \Psi(t),
\label{tkv2}
\end{equation}
where $f(r)$ is an arbitrary function on $r$ and 
\begin{equation}
\Psi=e^{\frac{1}{8} \int \alpha H_{T}dt}.
\label{tkv3}
\end{equation}
Then, reparameterizying $t$, holding the vector $\chi$ fixed  we may put $\Psi=1$, implying $H_T=0$, and thereof  that the asymmetry factor for our system is given by $H_{TF}$  alone.

Using the above result in (\ref{nd1})-(\ref{nd3}) we obtain 
\begin{equation}
A=f(r),\quad B=g(r)e^{\frac{\alpha}{3}\int H_{TF}dt},\quad
\alpha R=h(r)e^{-\frac{\alpha}{6}\int H_{TF}dt},
\label{nd12b}
\end{equation}
where $f, g, h$ are arbitrary dimensionless functions.

Using these  results  we shall build up  our two models corresponding to the adiabatic and the dissipative case respectively.
\section{Non dissipative case $q=0$.}
If we assume that the fluid is non-dissipative then using (\ref{nd12b}) in 
 (\ref{13})  (with $q=0$) produces 
\begin{equation}
f(r)=\gamma h^3(r),
\label{nd14}
\end{equation}
where $\gamma$ is a dimensionless constant.

Let us assume for $H_{TF}(t)$ the form
\begin{equation}
H_{TF}=\alpha(t-t_0),\qquad t_0=constant.
\label{nd21}
\end{equation}

Thus our system evolves without admitting a TKV, until $t=t_0$ when the system admits momentarily a TKV, which disappears afterward.

So far the models are defined up to two arbitrary functions of $r$ and some arbitrary constants. 
In order to specify a model, let us introduce the dimensionless  variables
\begin{equation}
x\equiv\frac{r}{r_\Sigma},\qquad y=\alpha(t-t_0)=H_{TF}.
\label{mond1}
\end{equation}
It is  worth emphasizing that the last equality in the above equation is valid only for the specific choice for $H_{TF}$ given by (\ref{nd21}). Indeed, $y$ is just a time-like coordinate whereas $H_{TF}$ is a geometrical variable endowed with a distinct geometrical meaning discussed before.

Next, we shall choose for the free parameters the value
\begin{equation}
\gamma= \alpha r_\Sigma=1.
\label{mond2}
\end{equation}

Finally the form of the functions $f(r)$ and $h(r)$ is assumed to be
\begin{equation}
f(r)=g(r)=h^3(r)=\alpha r=x.
\label{mond3}
\end{equation}

Under the above conditions the physical variables read
\begin{eqnarray}
\frac{8\pi \mu}{\alpha^2}=\frac{e^{-y^2/3}}{x^2}\left(\frac{1}{x^2}+x^{4/3} e^{y^2/2}\right)-\frac{y^2}{12 x^2},
\label{mund1b}
\end{eqnarray}

\begin{eqnarray}
\frac{8\pi P_r}{\alpha^2}=\frac{7e^{-y^2/3}}{9 x^4}-\frac{e^{y^2/6}}{x^{2/3}}-\frac{1}{3x^2}(\frac{y^2}{4}-1),
\label{mund2b}
\end{eqnarray}

\begin{eqnarray}
\frac{8\pi P_\bot}{\alpha^2}=-\frac{1}{6x^2}-\frac{y^2}{12x^2}-\frac{11e^{-y^2/3}}{9 x^4},
\label{mund3b}
\end{eqnarray}
while the expressions for the complexity factor,  the mass function, and the surface redshift  become

\begin{eqnarray}
\frac{Y_{TF}}{\alpha^2}=-\frac{4e^{-y^2/3}}{3x^4}-\frac{6+y^2}{12x^2},
\label{mund3b}
\end{eqnarray}

\begin{eqnarray}
m \alpha=\frac{x^{1/3} e^{-y^2/12}}{2}\left(\frac{y^2 e^{-y^2/6}}{36 x^{4/3}}-\frac{e^{-y^2/2}}{9x^{10/3}}+1\right),
\label{mund4b}
\end{eqnarray}
which evaluated at the boundary surface $x=1$ reads
\begin{eqnarray}
m _\Sigma\alpha=\frac{e^{-y^2/12}}{2}\left(\frac{y^2 e^{-y^2/6}}{36}-\frac{e^{-y^2/2}}{9}+1\right),
\label{mund45b}
\end{eqnarray}
and

\begin{equation}
z_\Sigma=\frac{6 e^{y^2/4}}{2-y e^{y^2/6}}-1.
\label{bhfnq}
\end{equation}

At this point the following remarks are in order
\begin{itemize}
\item As it follows from (\ref{mund1b})-(\ref{mund3b}), physical variables are singular at $r=0$. Accordingly the center of the distribution should be excluded from  consideration. However this fact should not bother us  since the expansion scalar vanishes and,  as it is known in such a case,  the center is surrounded by a vacuole inside which we assume vacuum \cite{exf}. Or, in other words, our fluid distribution is restricted to the region within the interval of $x$ $[x\ge x_{min}>0, 1]$ where $x_{min}$ corresponds to the boundary of the vacuole. In such a case, one should in principle impose matching conditions on the inner boundary surface. However as we shall  discuss in the last section, we shall not take into account the influence of these matching conditions in the dynamics of our models.
\item The Darmois conditions are not satisfied on the boundary surface  $r=r_\Sigma$. The radial pressure at the surface   vanishes only for $y\approx .47, -.47$.
\end{itemize}

In Figure 1, we have plotted the evolution of the complexity factor, the areal radius of the boundary surface, and  the radial pressure at the boundary surface, whereas in Figure 2 we have plotted  the anisotropic factor $8\pi(P_r-P_\bot)$, the mass function evaluated at the boundary, and the surface redshift. These  results will be discussed further in the last section.

\section{Dissipative case  $q\neq0$}
Let us now consider the dissipative case. Using the form (\ref{tkv1}) of the tensor $H_{\alpha \beta}$, and $H_{TF}=\alpha(t-t_0)$, we obtain as before 
\begin{equation}
A=f(r),\quad B=g(r)e^{y^2/6}\quad
\alpha R=h(r)e^{-y^2/12},
\label{nd12}
\end{equation}
where $f, g, h$ are arbitrary dimensionless functions.

In this case any possible model is determined up to three functions of $r$.

In order to specify our model we shall proceed as in the previous case. 

Thus, let us introduce the dimensionless  variables
\begin{equation}
x\equiv\frac{r}{r_\Sigma},\qquad y\equiv\alpha(t-t_0)\equiv H_{TF},
\label{mond1}
\end{equation}
as in the previous case  the above equation is valid only for the specific choice for $H_{TF}$ given above.

Next, we shall chose for the free parameters the value
\begin{equation}
\alpha r_\Sigma=1.
\label{mond2}
\end{equation}
\

Finally, to fully determine the model we need to specify the functions $f$, $h$ and  $g$. Since we assume $q\neq0$ we have to exclude a relationship like (\ref{nd14}), any other choice being (in principle) acceptable. 

Thus let assume
\begin{equation}
f(r)=g(r)=h(r)=\alpha r=x.
\label{deff}
\end{equation}
Under these conditions the physical variables read
\begin{eqnarray}
8\pi \mu/\alpha^2 =\frac{e^{-y^2/3}}{x^4}-\frac{y^2}{12x^2}+\frac{e^{y^2/6}}{x^2},
\label{muidscosf}
\end{eqnarray}

\begin{eqnarray}
8\pi q/\alpha^2=-\frac{2ye^{-y^2/6}}{3x^3},
\label{qdisb}
\end{eqnarray}

\begin{eqnarray}
8\pi P_r/\alpha^2=\frac{1}{x^2}\left(\frac{1}{3}-\frac{y^2}{12}+\frac{3e^{-y^2/3}}{x^2}-e^{y^2/6}\right),
\label{pridscosfb}
\end{eqnarray}

\begin{eqnarray}
8\pi P_\bot/\alpha^2=-\frac{1}{x^2}\left(\frac{1}{6}+\frac{y^2}{12}+\frac{e^{-y^2/3}}{x^2}\right),
\label{ptdscosfb}
\end{eqnarray}

Finally from (\ref{14a}) and (\ref{lum}) we obtain 
\begin{equation}
L_\Sigma=-\frac{y e^{-y^2/3}}{3},
\label{lums}
\end{equation}
and
\begin{eqnarray}
L_\infty=-\frac{ye^{-5y^2/6}}{3}\left(1+\frac{y^2 e^{y^2/3}}{36}-\frac{y e^{y^2/6}}{3}\right),
\label{lumsin} 
\end{eqnarray}
whereas the expressions for the complexity factor and the redshift read
\begin{equation}
Y_{TF}/\alpha^2=-\frac{2e^{-y^2/3}}{x^4}-\frac{1}{2x^2}-\frac{y^2}{12x^2},
\label{bhf}
\end{equation}
and 
\begin{equation}
z_\Sigma=\frac{6 e^{y^2/4}}{6-y e^{y^2/6}}-1.
\label{zdis}
\end{equation}
Finally, using  the truncated version of the transport equation (\ref{5trun}), and (\ref{qdisb}) we may obtain an expression for the temperature. It reads
\begin{eqnarray}
 T^\ast \equiv \tilde T-\frac{\tilde T_{\Sigma}}{x}=\frac{y \ln x}{12\pi x}+\frac{\tau^\ast \left(\frac{1}{x}-1\right)\left(y^2-3 \right)}{36\pi x},
\label{tdis}
\end{eqnarray}
where $\tau^\ast=\alpha \tilde \tau$  and $\tilde T=\frac{\kappa T}{\alpha}$. This choice of variable ($T^\ast$) allows us to skip the adoption of  a specific value for $\tilde T_\Sigma$, which within the context of this study is not available. It should be clear that when dealing with a specific astrophysical phenomenon, such an information must be provided.

To highlight the effects derived from the appearance of the TKV on the evolution of the fluid we have plotted the complexity factor, $L_\infty$, and $z_\Sigma$ in Figure 3. In Figure 4 we have plotted the temperature ($T^\ast$) for $\tau^\ast=0$ and $\tau^\ast=1$, and the anisotropic factor. All these results together with the ones obtained from the previous case will be discussed next.

\begin{figure}[h]
\begin{center}
\includegraphics[width=5.in,height=6.in,angle=0]{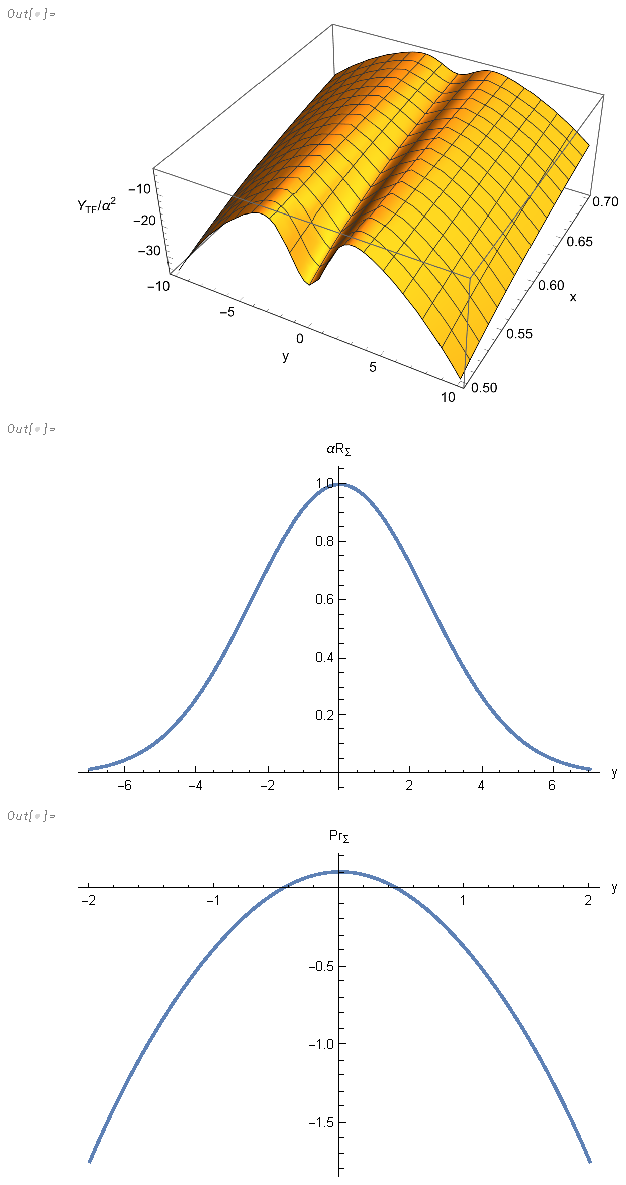}
\caption{$Y_{TF}/\alpha^2$  as function of $y$ and $x$  in the interval $(y,-10,10)$, $(x, .5, .7)$;  $\alpha R_\Sigma$ in the interval $(y, -7, 7)$; $P_{r\Sigma}$ in the interval $(y, -2, 2)$ for the non-dissipative case.}.\label{Figure 1}
\end{center}
\end{figure}

\begin{figure}[h]
\begin{center}
\includegraphics[width=5.in,height=6.in,angle=0]{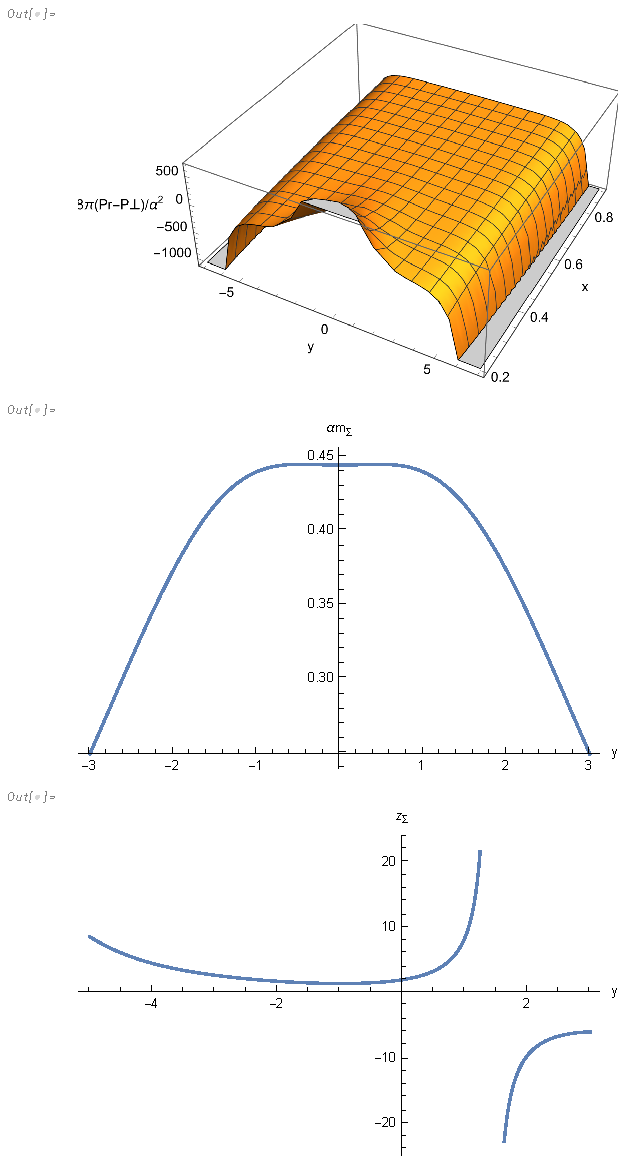}
\caption{$8\pi(P_r-P_\bot)/\alpha^2$ as function of $y$ and $x$  in the interval $(y, -7, 7)$, $(x, .2, .9)$; $\alpha m_\Sigma$  and $z_\Sigma$  in the interval $(y, -3,  3)$ and  $(y, -5,  3)$ respectively, for the non-dissipative case.}.\label{Figure 2}
\end{center}
\end{figure}

\begin{figure}[h]
\begin{center}
\includegraphics[width=5.in,height=6.in,angle=0]{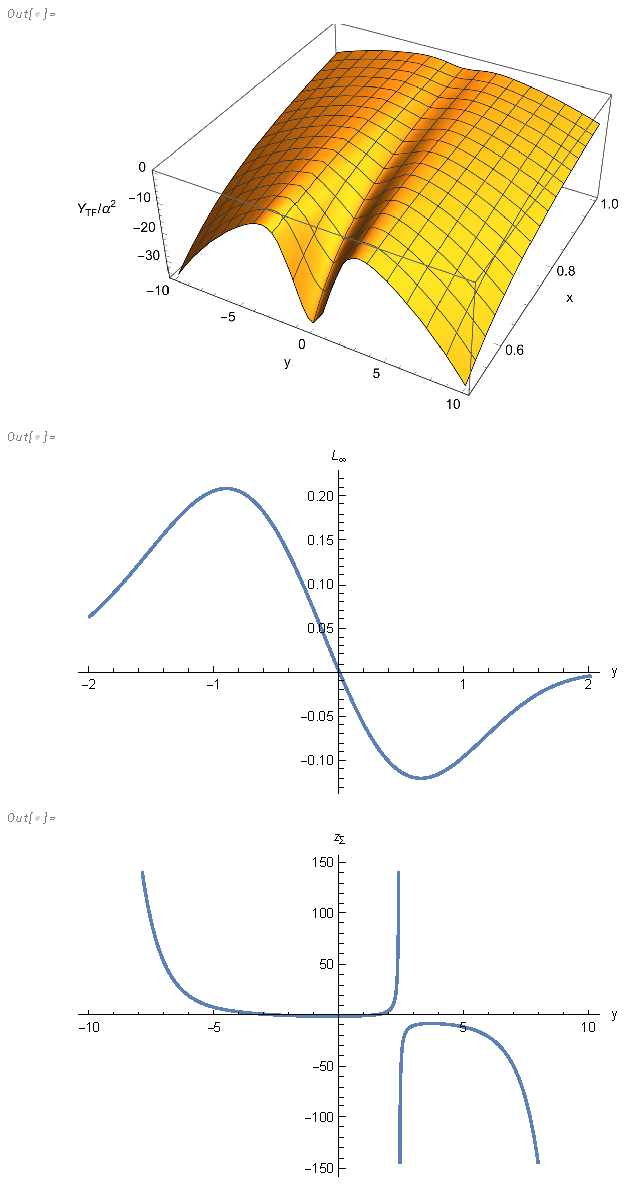}
\caption{$Y_{TF}/\alpha^2$  as function of $y$ and $x$  in the interval $(y,-10,10)$, $(x, .5,  1)$; $L_\infty$, and $z_\Sigma$  in the interval $(y, -2, 2)$,  and $(y, -10, 10)$ respectively, for the dissipative case}.\label{Figure 3}
\end{center}
\end{figure}
.
\begin{figure}[h]
\begin{center}
\includegraphics[width=5.in,height=6.in,angle=0]{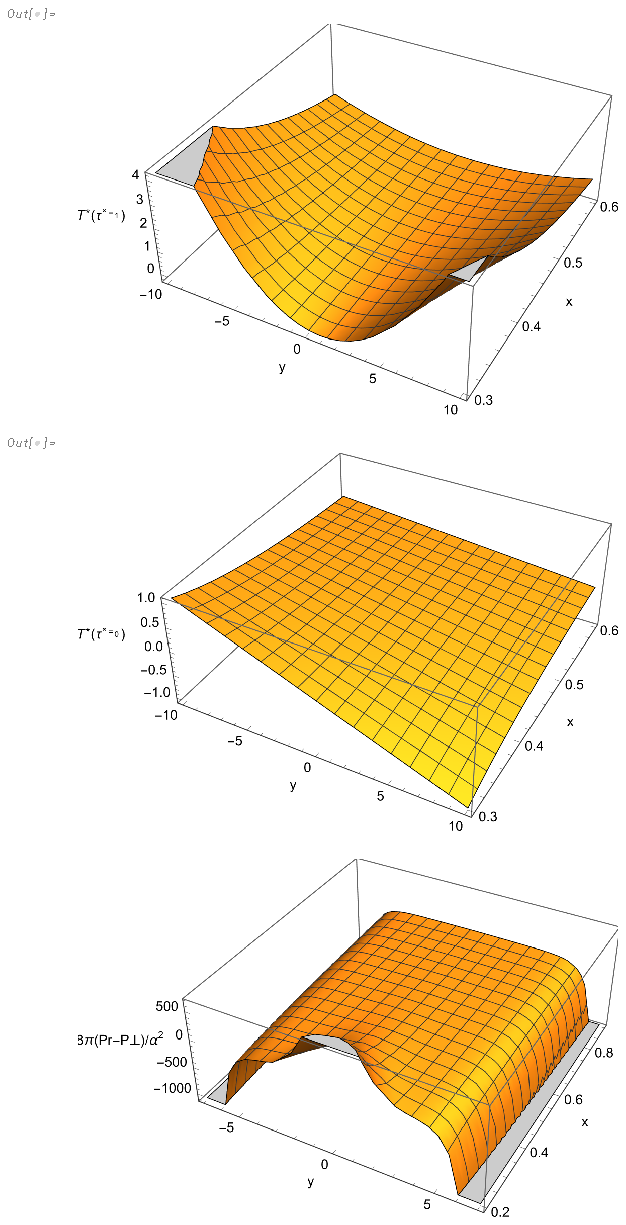}
\caption{$T^\ast$ as function of $y$ and $x$  in the interval $(y, -10, 10)$, $(x, .3,  .6)$ for $\tau^\ast=1$ and   $\tau^\ast=0$;  $8\pi(P_r-P_\bot)/\alpha^2$ as function of $y$ and $x$  in the interval $(y, -5, 5)$, $(x, .2, .7)$; for the dissipative case}.\label{Figure 4}
\end{center}
\end{figure}

\section{Discussion and Conclusion}
Our goal in this work is twofold. On the one hand we want to bring out the role played by the sudden appearance of a TKV,  on some physical aspects of evolving relativistic self-gravitating fluids. In particular  we are interested in relating the complexity factor to the instantaneous emergence of a TKV.    On the other hand we want to stress the potential relevance that  the approach based on the asymmetry factor may have in the modeling of the evolution of self-gravitating systems.

As mentioned before, the asymmetry factor  allows us to turn on (off) any type of symmetry at any given point of the evolution. A general formalism derived from the above mentioned concept was then applied to study two different models. It is worth emphasizing that these models are not intended to  describe any specific  astrophysical scenario,  but are presented with the purpose mentioned above.

The first model was assumed to be non-dissipative. It was obtained under the specific choice of the two arbitrary functions defining all adiabatic models admitting instantaneously a TKV. 
The choice adopted is justified by its simplicity. We shall refer to this issue at the end of this section.

Under the conditions above we obtain the model described by (\ref{mund1b})--(\ref{bhfnq}). As it is apparent from Figure 1, it represents an expanding fluid sphere for $y<0$ reaching a maximum at $y=0$ (the moment of appearance of the TKV), contracting from there on.

As indicated in the graphic for $P_{r\Sigma}$ of Figure 1, Darmois conditions are not satisfied on the boundary surface of the fluid distribution. This is confirmed by the discontinuity of the mass function across  the boundary surface  as shown in the graphic for $\alpha m_\Sigma$ in Figure 2. Accordingly we assume that the boundary is a  thin shell. All this having been said, it should be clear that the appearance of a thin shell implies that,  according to Israel approach \cite{isjc},   the boundary is endowed with some specific material properties (surface energy-density and tension) whose role in the analysis of our model has not been taken into account (neither for the outer nor for the inner boundary). It goes without saying that in a model intended to describe a specific astrophysical object, the dynamics of the shell might be a relevant factor and its role and the nature of its physical origin should be included in the discussion.
 
 There  are three remarkable  characteristics of this model, illustrated by Figures 1 and 2, which deserve to be emphasized
\begin{itemize}
\item The complexity factor presents a local minimum at $y=0$. While it is true that such stationary value depends on the specific form of $H_{TF}$, it is also true that 
the existence of a stationary value of the complexity factor at $y = 0$  appears also in the case of a conformal Killing vector (CKV) analyzed in
\cite{1}, thereby  suggesting that a possible model independent link might exist between complexity and 
symmetry.
\item The anisotropic factor    presents a local maximum  at $y=0$.
\item There is a discontinuity of the surface redshift around $y\approx 1.42$. Furthermore the sign of $z_{\Sigma}$ changes along the evolution (from blueshift to redshift). If such unexpected behavior is ever observed in astrophysics, then we could try to describe the evolution of the observed system through the asymmetry factor approach.
\end{itemize}

The second model is dissipative, it was obtained for the  choice of  the arbitrary functions given in (\ref{deff}). As in the previous case the above mentioned choice obeys to a criterion of   simplicity.  The full description of the model is provided by equations (\ref{muidscosf})--(\ref{tdis}).  As in the previous example, Darmois conditions are not satisfied on the boundary surface (the condition (\ref{j3}) is not satisfied identically).

Particular attention deserves the behavior of $T^\ast$ depicted in Figure 4. In the upper graphic $T^\ast$ is plotted for $\tau^\ast=1$, it was found that the same behavior (qualitatively) is observed for any value of $\tau^\ast \neq0$, namely:  $T^\ast$  decreases as $y$ approaches $0$ increasing afterward. However if we assume that the relaxation time vanishes then, as depicted in the corresponding graphic,  $T^\ast$ monotonically decreases after passing through $y=0$ (notice that, unlike the ``physical'' temperature $T$, the variable $T^\ast$ may be negative). In other words the increasing of  $T^\ast$ for $y>0$ is a relaxational effect, expected to be  observed when the value of the relaxation time is equal to, or larger  than, the characteristic  time of the evolution. 

Thus, in this specific model we are able to detect the appearance of thermal effect directly related to relaxation time and to the vanishing of the asymmetry factor. It is worth noticing that while the quantitative details of such an effect would depend on the transport equation employed, and the boundary temperature, the fact of its appearance, doesn't. 

At this point it is worth to recall that although
$\tau$ is very small for phonon--electron, and phonon--phonon interaction
at room temperature (${\cal O}(10^{-11})$ and ${\cal O}(10^{-13})$ seconds,
respectively \cite{Peierls}), neglecting it is the source of difficulties,
and in some cases a bad approximation. Thus, for example,  in superfluid Helium
\cite{pes,Peshkov} the relaxation time is of the order of $10^{-3}$ seconds.

Also, if the heat conductivity is dominated by electrons (as is the case of
a neutron star) \cite{Flowers1,Flowers2,Shibazaki}, the relaxation time may be in a  range as large as $[10^{-4}, 10^2]$ seconds, depending on the temperature of the object. Thus for the temperature proposed by Harwit \cite{Harwit} in the degenerate core of aged stars ($T\approx 10^7 K$) the
thermal relaxation time can be as high as $1$ second (see \cite{mdv, HP} and references therein for further discussion on this issue).

Next, as it follows from (\ref{nd12}), this model (as the previous one) represents an initially expanding sphere reaching a maximum value of $R_\Sigma$ at $y=0$ (when  the TKV appears) becoming a contracting sphere afterward.  The graphic of the luminosity at the boundary surface in Figure 3 indicates that it vanishes at $y=0$ changing from positive to negative for $y>0$.  As in the previous case the absolute value of the anisotropic factor, and the complexity factor have local minima at $y=0$ as depicted in Figures 2 and 3.  Also $z_\Sigma$ presents a discontinuity  (at $y\approx 2.36$) changing from redshift to blueshift during the evolution. 

From the above comments it is apparent that the instantaneous appearance of a TKV produces important effects on the subsequent evolution of the system.

These effects are of two kinds. On the one hand, from a purely theoretical point of view, the effects on the complexity factor  which bring out a link between  complexity and symmetry, the reach  of this fact deserves  further research.
On the other hand some of  the above mentioned effects  are directly related to observable variables (e.g luminosity and the gravitational redshift), this  paves the way  to explain, by resorting to a general approach based on  the asymmetry factor,
 some observational astrophysical or cosmological data, which appears unintelligible in terms of the standard theoretical setup.
 
We would like to conclude with the following final remarks 
\begin{itemize}
\item The asymmetry factor approach seems to be particularly suitable for describing  self-gravitating fluid distributions  exhibiting an unexpected behavior at some point of its evolution (e.g. a strange evolution of the redshift as the one exhibited in our models).
\item In any specific example, whenever it is possible,  we should distinguish between model dependent effects and effects associated exclusively to the instantaneous appearance of a symmetry. 
\item It is of utmost relevance  to find out if there is any physical phenomenon behind the appearance (disappearance) of any specific symmetry. In other words, we should  be able to identify a physical reason for the vanishing of the asymmetry factor.
\item The presented models were  not conceived to describe any specific physical  scenario, but just to bring out the effects produced by the sudden appearance of a TKV. Therefore neither the presence of a thin shell, nor the violation of some energy conditions are, at this point, relevant for the discussion. As mentioned  before in a model intended to describe a specific astrophysical object, the physical properties of a thin shell are relevant  and their  role and the nature of its physical origin cannot be overlooked. The same comment applies to any eventual violation of energy conditions.
\item As mentioned before the choice  for the asymmetry factor is justified by its simplicity. However  if we assume that $H_{TF}$ is  a well behaving  function, we may expand it in power series of $t-t_0$, implying that, at least, for values of $t$ not far from $t_0$ the above choice is likely to describe the evolution quite well. Notwithstanding, it is obvious that  $H_{TF}$ is a degree of freedom   of the system, and different choices may be invoked to describe different scenarios of physical interest. In the same line of arguments it is worth noticing that other forms of tensor $H_{\alpha \beta}$ may be proposed, which could involve more than one asymmetry factor.
\item Our models were obtained under the assumptions (\ref{mond3}) and (\ref{deff}) for the functions $f(r)$, $h(r)$ and $g(r)$, which as mentioned before were  adopted for simplicity.  In goes without saying that for the treatment of any specific astrophysical problem, the choice of these functions must be dictated by the very nature of the problem under consideration.
\item  While it is true that specific imprints of the instantaneous appearance of a TKV are clearly model dependent, the very fact that such an apearance drastically changes the subsequent evolution of the system, is not. Besides, for some variables (e.g., the complexity factor, and the anisotropic
factor) it seems to exist a persistent imprint for diferent models and types of symmetries.
\end{itemize}
\section*{Acknowledgments}
This  work  was partially supported by the Grant PID2024-158938NB-I00 funded by MICIU/AEI/ 10.13039/501100011033 and by ERDF A way of making Europe.

\end{document}